\documentclass[intlimits,twoside,a4paper]{article}
\usepackage{amsmath, amsfonts, amssymb}
\usepackage{euscript}
\usepackage{graphicx}
\usepackage[T2A]{fontenc}
\usepackage[cp1251]{inputenc}
\usepackage{epsf}
\usepackage{cite}
\usepackage{pstricks,pstricks-add}

\DeclareMathOperator{\arcosh}{arcosh}
\DeclareMathOperator{\rot}{rot}
\DeclareMathOperator{\sn}{sn}
\DeclareMathOperator{\dn}{dn}

\usepackage{cmpj2}

\issue{2016}{19}{3}{33604}
\doinumber{10.5488/CMP.19.33604}

\title[Equilibrium configurations of director]%
{Equilibrium configurations of director in a planar nematic cell with one spatially modulated surface}

\author[M.F. Ledney, O.S. Tarnavskyy, A.I. Lesiuk, V.Yu. Reshetnyak]{M.F. Ledney, O.S. Tarnavskyy, A.I. Lesiuk, V.Yu. Reshetnyak}
\address{Physics Department, Taras Shevchenko National University of Kyiv, Kyiv, Ukraine}

\authorcopyright{M.F. Ledney, O.S. Tarnavskyy, A.I. Lesiuk, V.Yu. Reshetnyak, 2016}
\date{Received February 18, 2016, in final form May 31, 2016}

\begin{document}

\maketitle

\begin{abstract}
We study two-dimensional equilibrium configurations of nematic liquid crystal (NLC) director in a cell confined between
two parallel surfaces: a planar surface and a spatially modulated one. The relief of the modulated surface is described by a smooth periodic sine-like function. The director easy axis orientation is homeotropic at one of the bounding surfaces and is planar at the other one.
Strong NLC anchoring with both surfaces is assumed.
We consider the case where disclination lines occur in the bulk of NLC strictly above local extrema of the modulated surface. These disclination lines run along the crests and troughs of the relief waves.
In the approximation of planar director deformations we obtain analytical expressions describing a director distribution in the bulk of the cell. Equilibrium distances from disclination lines to the modulated surface are calculated and their dependences on the cell thickness and the period and depth of the surface relief are studied.
It is shown that the distances from disclination lines to the modulated surface decrease as the depth of the relief increases.
\pacs 61.30.-v, 61.30.Hn, 61.30.Jf
\end{abstract}

\section{Introduction}

Over the recent years the number of researches in liquid crystals~(LCs) with the aim to design low power~LC displays with high resolution has been increasing very fast. In this field, bistable~LC systems have proved to be very promising.
A planar nematic liquid crystal (NLC) cell with at least one spatially modulated boundary surface that has the shape of a smooth or broken periodic line
is one of the well-known bistable~LC systems~\cite{Kriezis2001, Edwards2004, Ohzono2012, Dammone2012}.
Such systems are used in ZBD technology~\cite{Davidson2002, Spencer2006, Carbone2009, Spencer2010}, which allows one to create ergonomic display devices.
 While the cell of standard~LC display requires a constant power supply, bistable~LC systems only need the power to switch between different optic states caused by the existence of two or more stable director configurations. In the operation modes that do not require a frequent change of images, the employment of bistable LC systems as basic components of display devices is especially profitable since this reduces the total power consumption by such devices.

The modelling of display devices based on bistable~LC systems
aiming at finding the optimal operating mode requires
the knowledge of the director distribution in each stable state.
The director stable states between which the switching is performed
can have structure defects of disclination type caused by the relief surface,
flexopolarization, etc.

In most theoretical studies of NLC director configurations,
only semi-infinite NLC cells bounded by one spatially modulated surface were considered due to the complexity of calculations~\cite{Faetti_1987, Fournier_1999, Barbero_2008, Fukuda2D_2008, Fukuda_2008, Fukuda_2009, Davidson2012}.
 Many works studied the LC director distribution only in a thin NLC layer adjacent to the boundary surface. In particular, the peculiarities of~NLC elastic interaction with spatially modulated surface of semi-infinite NLC cell were studied in~\cite{Faetti_1987, Fournier_1999, Barbero_2008}. The interaction of NLC with the relief surface was studied theoretically in the case of two-dimensional spatial modulation of the surface~\cite{Fukuda2D_2008} and in the case of twisted LC director structure in the cell~\cite{Fukuda_2008, Fukuda_2009}.
The technique of the director structure calculation in semi-infinite NLC above the surface with the profile of periodically broken line was proposed in~\cite{Davidson2012}.
In~\cite{Ohzono2012}, possible stable LC director configurations
between a planar and a smooth spatially modulated substrates were experimentally studied.
Stable director states and switching between them were studied in an NLC cell with one grooved substrate~\cite{Uche2006}.
The authors of~\cite{Ladak2009, Evans2010} carried out numerical calculations of stable director configurations between two sawtooth substrates.
Two-dimensional director configurations in NLC cell with one sawtooth substrate were obtained~\cite{Romero-Enrique_2010, Rojas-Gomez_2012}, while taking into account the finiteness of the cell thickness.

In general, there are many studies on the director configuration states.
However, the finiteness of the cell thickness was considered only in some particular cases of sawtooth substrate relief.
Besides that, there is no analytical description of LC director configuration which would take into account the finiteness of cell thickness and the presence of defects in a cell with smooth periodic surface relief.
In the present paper, we propose a technique which allows one to calculate two-dimensional equilibrium configurations of the~NLC director in a planar nematic cell of finite thickness.
The cell is bounded by two surfaces one of which is planar and the other one is smooth and spatially modulated.
The anchoring of NLC with bounding surfaces is assumed to be strong.
If the orientation of director easy axis at one bounding surface is homeotropic and at the other one it is planar, then
disclination lines can occur in the bulk of NLC~\cite{Ohzono2012}.
In this paper we calculate equilibrium LC director configurations with the structure defects in the bulk.
The relief of the modulated surface is described by a smooth periodic sine-like function.
This function is not arbitrarily taken, but is chosen in a special way for the given values of the cell thickness, the period and depth of the relief.
We find equilibrium distances from disclination lines to the relief surface and study their dependence on the cell thickness and on the period and depth of the surface relief.
It is worth noting that the knowledge of the defects location is necessary in
nanoparticles trapping and arrangement by topological defect systems in LC cells~\cite{Yoshida, Coursault, Skarabot, Senyuk, Jose}.

\section{The problem geometry and basic equations}

\begin{figure}[!b]
\begin{center}
\includegraphics[width=0.65\textwidth]{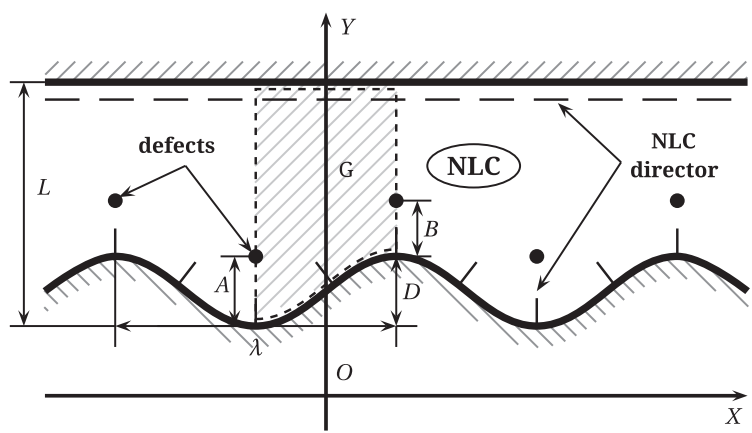}
\end{center}
\caption{The problem geometry: $L$~is the cell thickness, $\lambda$, $D$~are the spatial period and the depth of the modulated surface relief, $A$, $B$~are the distances from the disclination lines to the cell relief surface, G~is the region in which, for symmetry reasons, the angle $\theta(X,Y)$ between the director and $OX$-axis should be found.
}
\label{fig1}
\end{figure}

Let us consider a planar cell of~NLC which is bounded in the direction of the coordinate axis~$OY$ by two surfaces: a planar surface and spatially modulated one (see figure~\ref{fig1}). We assume that the shape of the modulated surface relief is described by a smooth sine-like function periodic in the coordinate $X$ with the
period~$\lambda$. The NLC anchoring energy is supposed to be infinitely large. The director easy axis at the planar surface is assumed to be directed along the axis~$OX$, while at the spatially modulated surface it is homeotropic, i.e., perpendicular to the surface.

We consider the case where under the stated conditions for the director at bounding surfaces, the director field in the bulk of the cell is not continuous. Therefore, the disclination lines with the strength ``$\pm 1/2$'' occur strictly above the local maxima and minima of the spatially modulated surface. These disclination lines are perpendicular
to~$XOY$ plane and, correspondingly, parallel to the relief waves (see figure~\ref{fig1}).
In a general case, disclination lines can adopt a zigzag form. Nevertheless, as was shown in~\cite{Ohzono2012_nc}, if elastic constants are equal $K_1=K_2=K_3=K$, then the disclination lines are straight.
Here and further on, we assume that the one elastic constant approximation holds.

\begin{figure}[!b]
\begin{center}
\includegraphics[width=0.4\textwidth]{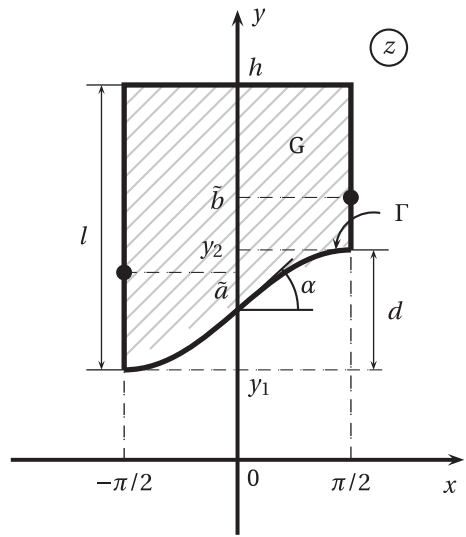}
\end{center}
\caption{The geometry of region~G in which the director angle $\theta(x,y)$ is sought as a solution of Laplace's equation: $\bullet$ --- disclination lines,
$y1$, $y2$ are distances from $Ox$-axis to a crest and a trough of the relief surface, respectively, $\tilde a$, $\tilde b$, $h$ are distances from the defects and from the upper cell surface to $Ox$-axis,
$\Gamma$ is the curve which describes the form of the relief surface,
$\alpha$ is the angle between the tangent to $\Gamma$ and the direction of $Ox$-axis,
$l=kL$, $d=kD$~are dimensionless thickness of the cell and the dimensionless depth of the modulated surface relief, $k=2\pi/\lambda$.}
\label{fig2}
\end{figure}

The deformations of the director field in the cell are assumed to be planar, that is, lying in the plane~$XOY$. Since the system is homogeneous in the direction of $OZ$-axis, we write the~NLC director in the form
\begin{equation}\label{n}
{\bf n}={\bf i}\cdot\cos\theta(X,Y)+{\bf j}\cdot\sin\theta(X,Y),
\end{equation}
where~$\bf i$, $\bf j$ are unit vectors of the axes of Cartesian coordinate system.
The twisting deformations (${\bf n}\cdot\rot {\bf n}=0$) are not involved in this case and the free energy of the~NLC cell is equal to $\displaystyle F=\dfrac{K}{2}\int\nolimits_V (\nabla\theta)^2\,\rd V$. Then, the director distribution in the bulk of NLC is described by
Lapalce's
equation~$\Delta_{X,Y} \theta=0$. For symmetry reasons, we seek a solution~$\theta(X,Y)$ of Laplace's equation in the region~G marked by a dashed line in figure~\ref{fig1}. It is evident that at the boundary of region~G, the unknown function~$\theta(x,y)$ should satisfy the following conditions (see figure~\ref{fig2}):
\begin{equation}\label{theta_S}
\theta_\text{S}=\left\{
\begin{array}{ll}
-\pi/2, & \quad\mbox{if}\qquad x=-\pi/2,\,\, y_1\leqslant y<\tilde a,\\
0, & \quad\mbox{if}\qquad x=-\pi/2,\,\, \tilde a< y\leqslant h,\\
0, & \quad\mbox{if}\qquad -\pi/2\leqslant x\leqslant\pi/2,\,\, y=h,\\
0, & \quad\mbox{if}\qquad x=\pi/2,\,\, \tilde b< y\leqslant h,\\
-\pi/2, & \quad\mbox{if}\qquad x=\pi/2,\,\, y_2\leqslant y<\tilde b,\\
\alpha-\pi/2, & \quad\mbox{if}\qquad (x,y)\in\Gamma,\\
\end{array}
\right.
\end{equation}
where~$x=kX$, $y=kY$~are dimensionless coordinates, $k=2\pi/\lambda$,
$\tilde a=y_1+a$, $\tilde b=y_2+b$~are dimensionless distances from the disclination lines to the coordinate plane~$xOz$,
$\alpha$~is the angle between a tangent line to the curve~$\Gamma$ at a given point and the positive direction of $Ox$-axis, $a=kA$, $b=kB$.

\begin{figure}[!t]
\begin{center}
\includegraphics[width=0.3\textwidth]{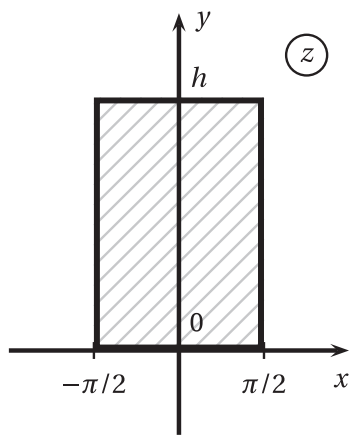}
\end{center}
\caption{The rectangle region of complex plane~$z$, which is conformally mapped on the upper half plane~$\Im  w\geqslant 0$ by function~(\ref{3}).}
\label{fig3}
\end{figure}

The function
\begin{equation}\label{3}
w=\sn \left(\dfrac{2}{\pi}K(m)z;\, m\right)
\end{equation}
is a conformal mapping which maps
the rectangle with the height~$h$ and the width~$\pi$ in the complex plane~$z=x+\ri y$ (see figure~\ref{fig3}) onto the upper half plane of the complex plane~$w$.
Here, $\sn (z;\,m)$ is the Jacobian elliptic function, $K(m)$ is the complete elliptic integral of the first type. The rectangle vertexes are mapped onto the points~$-1/m$, $-1$, $1$, $1/m$ of the real axis of complex plane~$w$, where parameter~$m$ is a root of the equation (see \cite{Ablowitz})
\begin{equation}\label{4}
\dfrac{K(m)}{K\left(\sqrt{1-m^2}\right)}=\dfrac{\pi}{2h}\,.
\end{equation}

\begin{figure}[!b]
\begin{center}
\includegraphics[width=0.65\textwidth]{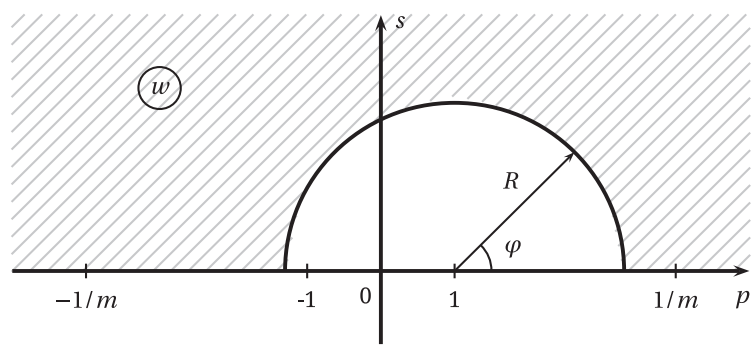}
\end{center}
\caption{The hatched region of complex plane~$w$ is conformally mapped on the upper half plane~$\Im  \eta\geqslant 0$ by function~(\ref{6}), while on the other hand it is mapped on the region~G in figure~\ref{fig2} by the function which is inverse to~(\ref{3}).}
\label{fig4}
\end{figure}

Let us cut out a semi-circle of radius~$R$ centred at point~$w=1$ of the real axis from the upper half plane~$\Im  w\geqslant 0$ (see figure~\ref{fig4}), where
\begin{equation}\label{5}
2\leqslant R\leqslant\dfrac{1}{m}-1.
\end{equation}
This obviously implies that
\begin{equation}\label{5a}
0\leqslant m\leqslant \dfrac{1}{3}\,.
\end{equation}

The upper half plane with the cut out semi-circle can be conformally mapped onto the upper half plane of complex plane~$\eta$ using the Zhukovsky transformation~\cite{Ablowitz}
\begin{equation}\label{6}
\eta(w)=\dfrac{1}{2}\left(\dfrac{R}{w-1}+\dfrac{w-1}{R} \right).
\end{equation}
On the other hand, the transformation~$z=\pi F(w;\, m)/[2K(m)]$, which is inverse to~(\ref{3}), maps the half plane $\Im  w\geqslant 0$ with the cut out semi-circle of radius~$R$ onto the rectangle of the complex plane~$\Im  z\geqslant 0$ (region~G in figure~\ref{fig2}), which is clipped from below by a certain smooth curve~$\Gamma$. Here, $F(w;\, m)$ is the elliptic integral of the first type. The curve~$\Gamma$ is the image of semi-circle~$w=1+R\re^{\ri\varphi}$, where~$\varphi\in[\pi,0]$. The tangent lines to curve~$\Gamma$ at points with~$x=\pm\pi/2$ prove to be parallel to~$Ox$-axis due to the principle of angle-preservation in conformal mapping~\cite{Ablowitz}. Combining functions~(\ref{3}) and~(\ref{6}) we obtain the following transformation
\begin{equation}\label{7}
\eta(z)=\dfrac{1}{2}\left[\dfrac{R}{\sn \bigl(2K(m)z/\pi; m\bigr)-1}+
\dfrac{\sn \bigl(2K(m)z/\pi; m\bigr)-1}{R} \right],
\end{equation}
which maps the region~G belonging to the complex plane~$z$ onto the upper half plane of the complex plane~$\eta$. Here, the lower boundary of~G which is the smooth curve~$\Gamma$ is given in the form
\begin{equation}\label{8}
z=\dfrac{\pi}{2K(m)}F\bigl(1+R \re^{\ri\varphi}; m\bigr), \qquad \mbox{where} \qquad \varphi\in[\pi,0].
\end{equation}

\section{The function of the surface relief}

Let us find a real function describing the relief of the cell modulated surface (curve~$\Gamma$ in figure~\ref{fig2}). Since function~(\ref{6}) maps the semi-circle~$w=1+R\re^{\ri\varphi}$, where~$\varphi\in[\pi,0]$, onto the real axis interval~$[-1,1]$ of the complex plane~$\eta=p+\ri s$, formula~(\ref{8}) can be rewritten as follows:
\begin{equation}\label{9}
x(p)+\ri y(p)=f(p)=\dfrac{\pi}{2K(m)}F\left(1+R \Big(p+\ri \sqrt{1-p^2}\Big);\, m\right),
\end{equation}
where~$-1\leqslant p\leqslant 1$.

Differentiating~(\ref{9}) with respect to the variable~$p$ and substituting~$p=\cos\varphi$ (since~$-1\leqslant p\leqslant 1$ for the points of curve~$\Gamma$), we obtain
\begin{equation}\label{13}
\begin{split}
& x'(\varphi)= -e^{\beta(\varphi)}\cos\alpha(\varphi)\,\sin\varphi,\\
& y'(\varphi)= -e^{\beta(\varphi)}\sin\alpha(\varphi)\,\sin\varphi,
\end{split}
\end{equation}
where~$\alpha=\Im  \ln f'(p)$, $\beta=\Re  \ln f'(p)$.
A prime is used to mark the derivatives with respect to the argument.

Substituting~$p=\cos\varphi$ into $f'(p)$ and expanding the functions of~$\ln (1+x)$ type at~$|x|<1$ into Maclaurin series, we find:
\begin{equation}\label{15}
\beta(\varphi)= \ln\left(\dfrac{\pi}{2K(m)}\,\dfrac{1}{\sqrt{1-m^2}}\,\dfrac{1}{\sin\varphi} \right)+\gamma(\varphi),\\
\end{equation}
\begin{equation}\label{16}
\alpha(\varphi)= \dfrac{1}{2}\sum_{k=1}^\infty \dfrac{(-1)^{k+1}}{k} \left[\left(\dfrac{2}{R}\right)^k-
\left(\dfrac{mR}{m-1}\right)^k-\left(\dfrac{mR}{1+m}\right)^k \right]\sin k\varphi,
\end{equation}
where
\begin{equation*}
\gamma(\varphi)= -\dfrac{1}{2}\sum_{k=1}^\infty \dfrac{(-1)^{k+1}}{k} \left[\left(\dfrac{2}{R}\right)^k+
\left(\dfrac{mR}{m-1}\right)^k+\left(\dfrac{mR}{1+m}\right)^k \right]\cos k\varphi.
\end{equation*}

From equations~(\ref{13}), taking into account the explicit form of~$\beta(\varphi)$~(\ref{15}), we obtain the expression for the curve~$\Gamma$ in parametric form
\begin{equation}\label{17}
\begin{split}
x(\varphi)= & -\dfrac{\pi}{2}+\dfrac{\pi}{2K(m)}\,\dfrac{1}{\sqrt{1-m^2}} \int\limits_\varphi^\pi \re^{\gamma(\varphi)} \cos\alpha(\varphi)\,\rd\varphi,\\
y(\varphi)= & y_1+\dfrac{\pi}{2K(m)}\,\dfrac{1}{\sqrt{1-m^2}} \int\limits_\varphi^\pi \re^{\gamma(\varphi)} \sin\alpha(\varphi)\,\rd\varphi,\\
\end{split}
\end{equation}
where~$\varphi\in[\pi,0]$. Formulae (\ref{17}) take into account that~$\varphi=\pi$ corresponds to the beginning of curve~$\Gamma$ at the
point~$(-\pi/2, y_1)$ (see figure~\ref{fig2}). $y$-coordinates of the ends of curve~$\Gamma$ can be obtained by taking the imaginary part of formula (\ref{8}) at~$\varphi=\pi,\, 0$:
\begin{equation}\label{19}
\begin{split}
y_1= & \dfrac{\pi}{2K(m)}\,F \left( \dfrac{1}{\sqrt{1-m^2}} \sqrt{1-\dfrac{1}{(R-1)^2}};\, \sqrt{1-m^2} \right),\\
y_2= & \dfrac{\pi}{2K(m)}\,F \left( \dfrac{1}{\sqrt{1-m^2}} \sqrt{1-\dfrac{1}{(R+1)^2}};\, \sqrt{1-m^2} \right).
\end{split}
\end{equation}

\begin{figure}[!b]
\vspace{-1mm}
\begin{center}
\includegraphics[width=0.65\textwidth]{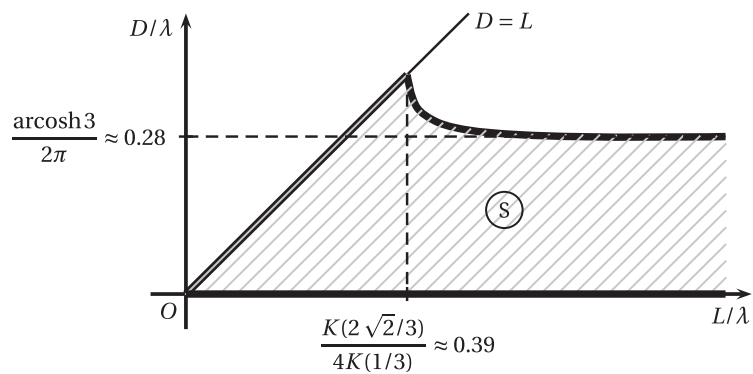}
\end{center}
\vspace{-2mm}
\caption{
The hatched region~S corresponds to admissible values of the cell thickness~$L$, the surface relief depth~$D$ and period~$\lambda$ for which a director distribution and equilibrium locations of the defects can be calculated.
}\label{region}
\end{figure}

Therefore, the relief of the modulated surface is described by the curve $\Gamma$ and, as follows from~(\ref{17}) and~(\ref{19}), is determined by the values of two dimensionless parameters~$m$ and~$R$. For the given values of geometric parameters of the NLC cell, namely, the cell thickness~$L$, the depth~$D$ and period~$\lambda$ of the modulated surface relief, the values of parameters~$m$ and~$R$ are found from the following system of equations
\begin{equation}\label{21}
\left\{
\begin{array}{l}
h-y_1=\dfrac{2\pi}{\lambda}\,L,\\[9pt] y_2-y_1=\dfrac{2\pi}{\lambda}\,D,
\end{array}\right.
\end{equation}
where~(\ref{4}) and~(\ref{19}) must be taken into account.
It is necessary that the obtained parameters $m$ and $R$ should meet
physically justifiable conditions~(\ref{5}) and (\ref{5a}).
Thus, $R = 1/m-1$ [see condition (\ref{5})] corresponds to the case $L=D$ where the upper planar surface lies on the lower relief surface of the cell.
If $R<2$, then the curve $\Gamma$, which describes the surface relief (see figure \ref{fig2}), becomes a broken line rather than a smooth curve.
It is worth noting that for $m>1/3$, inequality (\ref{5}) cannot be satisfied at all.
Conditions~(\ref{5}), (\ref{5a}) define the region of admissible values of $L$, $D$, $\lambda$.
This region which is found using (\ref{21}) is presented in figure~\ref{region}.
It is obvious that physically admissible values of the cell thickness $L$ and relief depth $D$ should meet the condition $0\leqslant D\leqslant L$.
Our conformal mapping technique is capable of describing the director field not for all $L$, $D$, $\lambda$, but only for those values of the parameters that correspond to the points of the region S.
Further on we consider only such values of parameters $L$, $D$, $\lambda$ from region S in figure~\ref{region} for which there exists a unique solution of system~(\ref{21}).

Let us consider a particular case $m=0$ which corresponds to a semi-infinite cell ($L\to\infty$).
Then, from~(\ref{19}) we have $y_1=\arcosh (R-1)$, $y_2=\arcosh (R+1)$.
Dimensionless relief depth $d=y_2-y_1$ reaches the maximum possible value~$d_{\text{max}}=\arcosh 3$ at~$R=2$ (see figure~\ref{region}).
At a fixed but still large parameter~$R\gg1$ ($h\to\infty$, and the value of~$y_1$~is finite), taking into account~(\ref{15}) and~(\ref{16}), from~(\ref{17}) we have $x(\varphi)\sim\pi/2-\varphi$ and $y(\varphi)\sim y_1+1/R+\cos\varphi/R$, where
$y_1\sim\ln(2R)-1/R$.
Hence, the function describing the curve~$\Gamma$ can be easily obtained in an explicit form $y(x)=\ln(2R)+\sin x/R$.
So, $2/R$ is the depth of the surface relief.
Thus, in a semi-infinite cell with a small surface relief depth, the form of the relief is described by function $\sin(x)$.

\section{The director field}

Now we find the director field~$\theta(x,y)$ in the region~G (see figure~\ref{fig2}).
By conformal mapping~$\eta(z)$~(\ref{7}), the boundary points of region G in the complex plane~$z$ are mapped onto the points of the real axis of the complex
plane~$\eta=p+\ri s$. Points~$p_1$ and~$p_2$ of the real axis~$p$ are the images of the defect points in the region~G, where~$-1/m\leqslant p_1 \leqslant -1$, $1\leqslant p_2\leqslant 1/m$. The conformal mapping of the region~G boundaries taking into account boundary conditions~(\ref{theta_S}) for function~$\theta(x,y)$ yields the following values of function~$\theta(p,s)$ on the real axis $p$:
\begin{equation}\label{gran-usl}
\theta(p,0)=\left\{
\begin{array}{ll}
0, & \qquad\mbox{if } \qquad -\infty<p<p_1\,, \,\, p_2<p<+\infty,\\[9pt]
-\dfrac{\pi}{2}\,, & \qquad\mbox{if } \qquad p_1< p<-1, \,\, 1< p<p_2\,,\\[9pt]
-\dfrac{\pi}{2}+\alpha, & \qquad\mbox{if } \qquad  -1\leqslant p\leqslant 1,
\end{array}\right.
\end{equation}
where~$\alpha$ is the angle between the tangent line to the curve~$\Gamma$
at a point~$z=x+\ri y$ and the positive direction of~$Ox$-axis. As follows from the properties of conformal mapping~\cite{Ablowitz}, the angle~$\alpha=\arg f'(p)=\Im  \ln f'(p)$, which is given by (\ref{16}) [a point~$z=f(p)$ belongs to the curve~$\Gamma$ at~$-1\leqslant p\leqslant 1$]. Therefore, in order to find the harmonic function~$\theta(\eta)=\theta(p,s)$ we have to solve the Dirichlet problem for the upper half plane~$\Im \eta\geqslant 0$.
This can be done by using the Poisson integral~\cite{Ablowitz}:
$\displaystyle\theta(\eta)=\dfrac{1}{\pi}\int\nolimits_{-\infty}^\infty \dfrac{s\theta(t,0)\, \rd t}{(p-t)^2+s^2}\,$. Substituting the value of~$\theta(p, 0)$~(\ref{gran-usl}) at the real axis~$\Re \eta$ into the integral, we obtain
\begin{equation}\label{Puasson}
\theta(\eta)=-\dfrac{1}{2}\int\limits_{p_1}^{p_2} \dfrac{s\, \rd t}{(p-t)^2+s^2}+
\dfrac{1}{\pi}\int\limits_{-1}^1 \dfrac{s\alpha(t)\, \rd t}{(p-t)^2+s^2}\,.
\end{equation}
After the substitution $t=\cos\varphi$, we use the value of~$\alpha(\varphi)$ taken from~(\ref{16}) in the second integral. As a result, the integration yields
\begin{equation}\label{27}
\theta(\eta)=  \Im \left\{\ln\sqrt{\dfrac{\eta-p_1}{\eta-p_2}}+\dfrac{1}{2}\sum_{k=1}^\infty \dfrac{(-1)^{k}}{k} \left[\left(\dfrac{2}{R}\right)^k-\left(\dfrac{mR}{m-1}\right)^k-\left(\dfrac{mR}{1+m}\right)^k \right]
\left(\eta-\sqrt{\eta^2-1} \right)^k\right\}.
\end{equation}

Substituting $\eta$ from formula~(\ref{7}) into~(\ref{27}) and reducing power series, after simple transformations, we obtain the following expression
\begin{equation}\label{29}
\theta(z)=\Im \ln\sqrt{\dfrac{1-2\xi(\tilde a)g(z)+g^2(z)}{1+2\xi(\tilde b)g(z)+g^2(z)}
\dfrac{\left[1-\frac{mR}{1-m}g(z)\right]\left[1+\frac{mR}{1+m}g(z)\right]}
{1+\frac{2}{R}g(z)}}\,,
\end{equation}
where
\begin{equation*}\label{30}
\xi(t)=-\dfrac{1}{2}\left[   
\dfrac{R}{\dn ^{-1}\left(\frac{2}{\pi}K(m)t; \sqrt{1-m^2} \right)-1}+
\dfrac{\dn ^{-1}\left(\frac{2}{\pi}K(m)t; \sqrt{1-m^2} \right)-1}{R}
\right],
\end{equation*}
\begin{equation*}
g(z)=\eta-\sqrt{\eta^2-1}=R\left[\sn \left(\dfrac{2}{\pi}K(m)z;\, m \right)-1\right]^{-1}.
\end{equation*}
Here, the function~$\dn (z;\, k)$ is the delta amplitude.

\section{The equilibrium positions of defects}

The equilibrium distances~$\tilde a$ and~$\tilde b$ from disclination lines to the coordinate plane~$xOz$ should provide the minimum value of the NLC free energy. By applying Green's first identity, the~NLC free energy of region~G per unit length along $Oz$-axis (see figure~\ref{fig2}) can be expressed in the form of a closed contour integral
\begin{equation}\label{31}
F=\dfrac{K}{2}\int\limits_{\, \text{G}}\!\!\!\!\int (\nabla\theta)^2\, \rd x \rd y=\dfrac{K}{2}\oint\limits_{\partial G}\theta
\dfrac{\partial\theta}{\partial n}\,\rd l,
\end{equation}
where~$\partial/\partial n$ is the normal derivative taken at the points of curve~$\partial G$, which restricts the region~G,
$\rd l$ is an arc element of the curve~$\partial G$. It is obvious that points~$(-\pi/2,\tilde a)$ and~$(\pi/2,\tilde b)$, which are points corresponding to defects, lie at the
curve~$\partial G$ and are the singular points of the free energy integral~(\ref{31}). For this reason, the integral~(\ref{31}) is calculated along the boundary of the region~G without two small areas around the defect points. After parametrization of the curve~$\partial G$ taking into account the bypass of the defect points, the free energy functional is further numerically minimized with respect to the positions of the defects.

\begin{figure}[!t]
\hspace*{40mm}\epsfysize=60mm  \epsffile{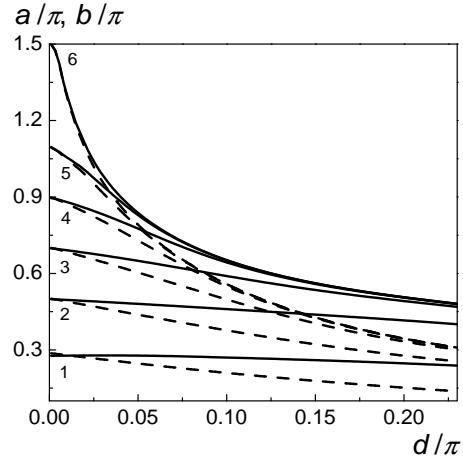}
\caption{Equilibrium distances $a$ (solid curves) and $b$ (dashed curves) from  disclination lines to the cell modulated surface versus the surface relief depth~$d$.
$l/\pi$=0.58\, (1); 1.0\, (2); 1.4\, (3); 1.8\, (4); 2.2\, (5); 2.6\,(6); 3.0\,(7).}
\label{fig5}
\end{figure}

Equilibrium distances from disclination lines to the cell modulated surface, $a=kA$ and~$b=kB$, as functions of dimensionless depth of the surface relief~$d=kD$ are calculated for several values of the cell dimensionless thickness~$l=kL$. The corresponding curves are presented in figure~\ref{fig5}. The distances are measured in the units of the region~G width which is equal to~$\pi$ (see figure~\ref{fig2}). As the surface relief depth~$d$ increases, the values of equilibrium distances $a$ and~$b$ decrease monotonously, but the difference between them~$|a-b|$ grows. Thus, the distance from the modulated surface to a disclination line located above a relief crest decreases faster than the distance to a disclination line located above a relief trough as the relief depth grows.

\begin{figure}[!b]
\vspace{-2mm}
\hspace*{40mm}\epsfxsize=60mm  \epsffile{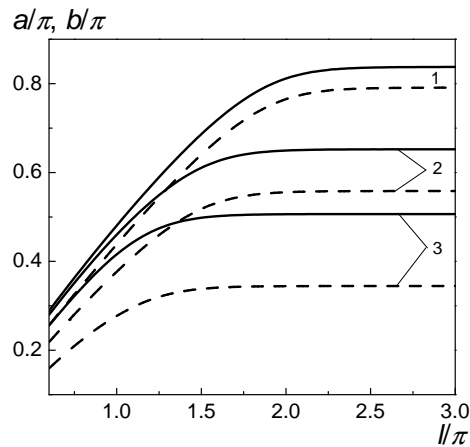}
\caption{Equilibrium distances $a$ (solid curves) and $b$ (dashed curves) from  disclination lines to the cell modulated surface versus the cell thickness~$l$. $d/\pi$= 0.05\, (1); 0.1\, (2); 0.2\, (3). }
\label{fig6}
\end{figure}

Equilibrium distances~$a$ and~$b$ from disclination lines to the cell modulated surface, calculated as functions of the cell thickness~$l$ for several values of the relief depth~$d$ are given in figure~\ref{fig6}. As the cell thickness~$l$ grows, the distances~$a$ and~$b$ increase monotonously until they reach certain constant values. Thus, for a given surface relief depth~$d$, the positions of defects become practically independent of the cell thickness when latter is sufficiently large. The greater the relief depth~$d$ of the modulated surface is, the faster the equilibrium distances~$a$ and~$b$ reach saturation.
It is worth noting that, for a given relief depth $d$, the form of the cell surface relief is not invariant with respect to the changes in the cell thickness~$l$.
This is just a peculiarity of the technique used by us.
However, this variation of the relief proves to be small, and we do not take it into account.
In particular, as calculations show, for the relief depth $d/\pi=0.2$,
the change in $l$ with the range as in figure \ref{fig6} corresponds to a relief variation that does not exceed 6 percent of the relief depth $d$.
Since for a given relief depth $d$, the defect locations are independent of the cell thickness $l$ starting from its certain value $l_\text{c}\lesssim\pi$ (see figure~\ref{fig6}), it follows that the approximation $m=0$
can be justifiably used in the case of a thick cell where~$l=kL\gg1$.
If the relief depth of a thick cell is small~$d=kD\lesssim1$ (while~$L\gg D$),
then the relief of the modulated surface can be described sufficiently well by the function~$y(x)\sim \sin x/R$.
Here, $R$ is found from system~(\ref{21}).
Therefore, in the case of cells with~$L\gg \lambda$, the
proposed technique allows one to calculate the director field and equilibrium locations of the defects, if the relief depth~$D<\lambda\,\arcosh 3/(2\pi)\approx 0.28 \lambda$.

\begin{figure}[!t]
\vspace{-1mm}
\hspace*{0mm}\epsfxsize=65mm  \epsffile{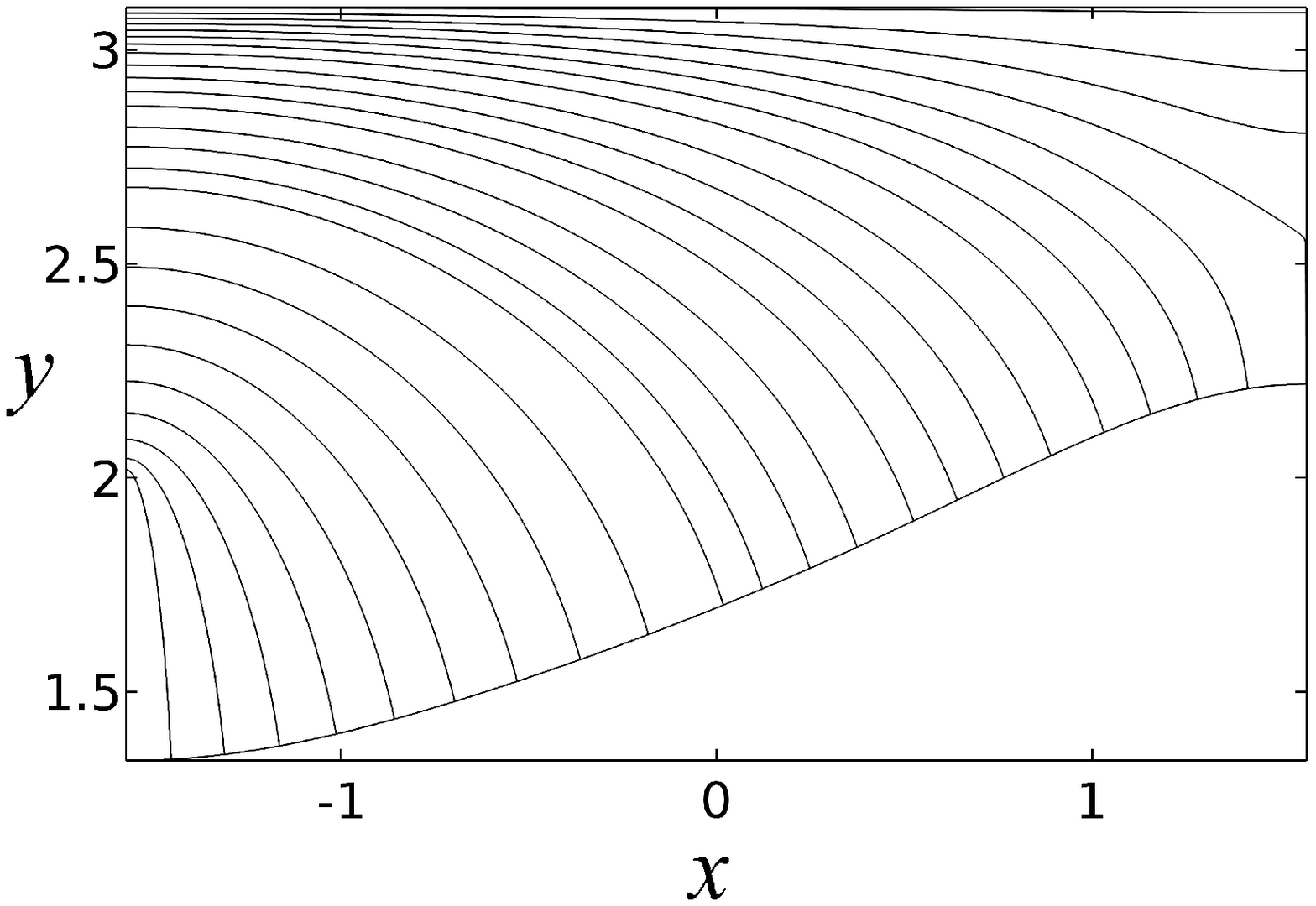}
\hspace*{5mm}\epsfxsize=70mm  \epsffile{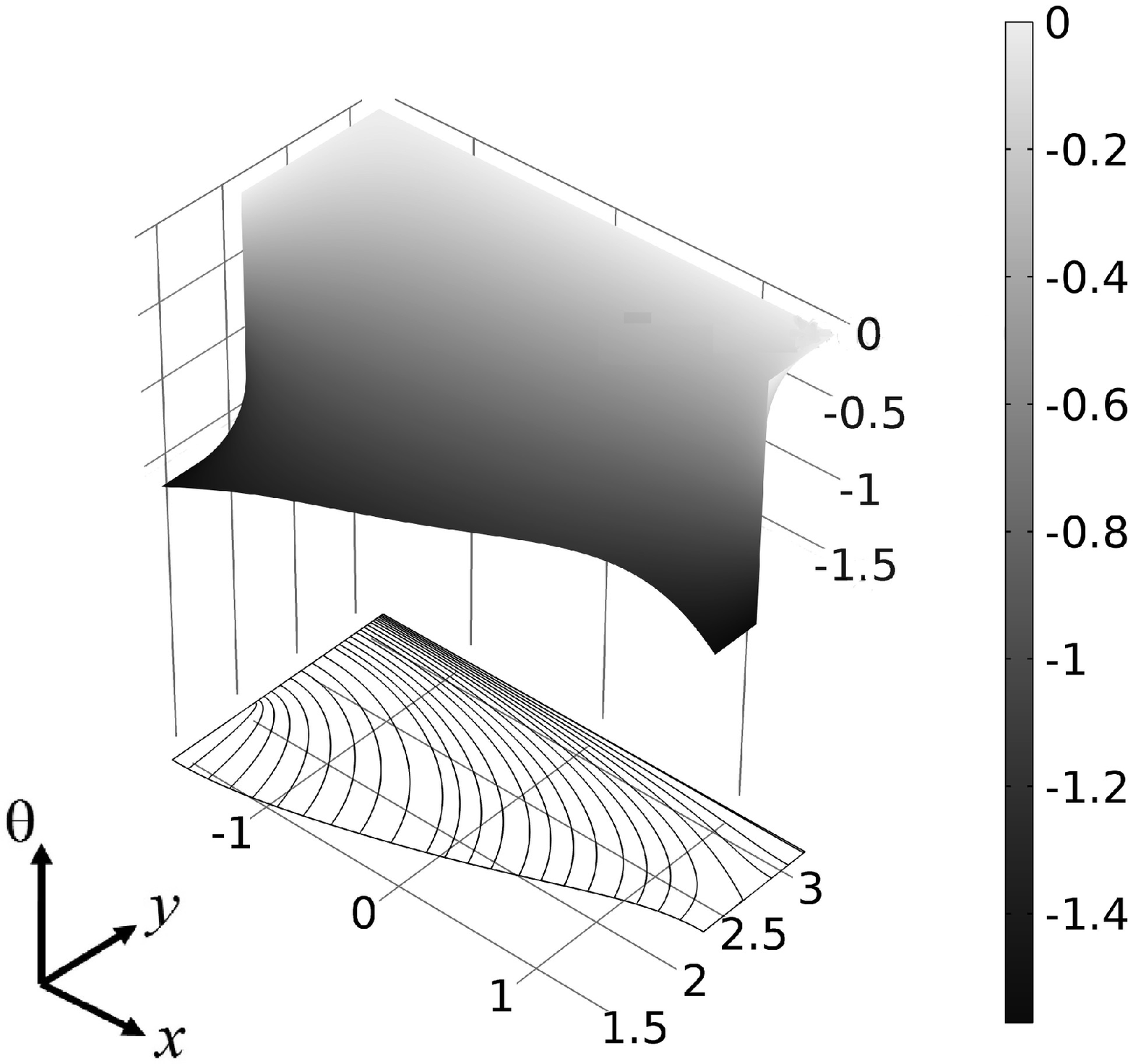}\\
\hspace*{30mm} (a) \hspace*{65mm} (b)
\caption{Director field lines~(a) and the distribution of the director deviation angle~$\theta$~(b) in the region~G.}\label{theta}
\end{figure}

In figure~\ref{theta}~(a), director field lines in the region~G are presented. The following~NLC cell parameters were used for calculations: $L=2.8~\mu \text{m}$, $\lambda=10~\mu \text{m}$ and~$D=1.4~\mu \text{m}$, which are close to the typical ones~\cite{Ohzono2012} and correspond to $m = 0.1789$ and $R = 2.9894$. The distribution of the director deviation angle~$\theta(x,y)$ in the region~G for the given cell parameters is presented in figure~\ref{theta}~(b).

Now we assume that the~NLC anchoring at the cell planar surface is homeotropic and strong. The~NLC anchoring at the spatially modulated surface is assumed to be planar: the director easy axis is tangent to the surface and lies in the plane~$XOY$.
Similar to the above considered case, disclination lines parallel to the relief wave crests and troughs occur in the bulk of NLC above the local extrema of the modulated surface. It is easy to see that the value of angle~$\theta$ in the region G is obtained by adding~$\pi/2$ to the value given by formula~(\ref{29}). It is evident that in this case for a given cell thickness~$L$, as well as depth~$D$ and period~$\lambda$ of the cell surface relief, the distances from disclination lines to the surface prove to be the same as those in the case of planar orientation of the director at the cell planar surface and homeotropic orientation of the director at the modulated surface.

\section{Conclusions}
\vspace{-1mm}

We have studied two-dimensional equilibrium configurations of the~NLC director in the cell bounded by two parallel surfaces: a planar surface and a spatially modulated one.
The NLC director anchoring with cell surfaces is assumed to be strong.
We have considered the case where the disclination lines occur in the bulk of NLC above local extrema of the modulated surface relief.
These disclination lines, which are caused by different orientation of the director easy axis at two bounding surfaces, namely, homeotropic at one surface and planar at the other one, run along the crests and troughs of the relief.
The relief of the modulated surface is described by a smooth sine-like periodic function.
This function is taken in a special way for particular measurable parameters of the cell, namely, the cell thickness~$L$, period~$\lambda$ and depth~$D$ of the surface relief.
Thus, for the given values of $L$, $\lambda$, $D$ system~(\ref{21}) together with~(\ref{4}) and~(\ref{19}) yields the values of~$m$ and~$R$. For the obtained values of~$m$ and~$R$, formulae~(\ref{17}) define a sine-like relief profile.
In particular, for a thick cell $L\gg \lambda$ with a small relief depth~$D\lesssim \lambda$, the relief profile is described by the function~$\sin x/R$.
The region of the admissible values of cell parameters $L$, $\lambda$, $D$
has been found, for which the proposed technique allows one to calculate a director distribution and equilibrium locations of the defects (see figure~\ref{region}).

In the approximation of planar director deformations, an analytically calculated director field is given by~(\ref{29}). Minimizing the free energy~(\ref{31}) of the system we have found equilibrium distances~$a=2\pi A/\lambda$
and~$b=2\pi B/\lambda$ from disclination lines to the cell relief surface.
We have studied the dependence of these distances on the cell thickness~$L$ and the period~$\lambda$ and depth~$D$ of the surface relief.
The distances from disclination lines to the cell modulated surface decrease as the relief depth increases (see figure~\ref{fig5}), whereas the distance between the modulated surface and the disclination line located above a relief wave crest decreases faster than the distance corresponding to the disclination line located above a relief trough.
The increase in the cell thickness results in greater equilibrium distances
from disclination lines to the modulated surface. However, they become practically independent of the thickness for sufficiently large thickness value (see figure~\ref{fig6}).
The presented in figures~\ref{fig5}, \ref{fig6} equilibrium distances~$a$ and~$b$ from disclination lines to the relief surface can be used in constructing NLC cells, one surface of which has a sine-like relief profile.

The knowledge of the defects location and its dependence on the cell parameters
can be useful for investigations in trapping and self-assembling of nanoparticles by topological defects in NLC cells.

\section*{Acknowledgements}

The authors express sincere gratitude to I.P.~Pinkevich for helpful advice during the result discussions.

\ukrainianpart

\title{Рівноважні конфігурації директора в планарній нематичній
комірці з однією просторово модульованою поверхнею}
\author{М.Ф. Ледней, О.С. Тарнавський, А.І. Лесюк, В.Ю. Решетняк}
\address{Київський національний університет імені Тараса Шевченка,
Київ, Україна
}

\makeukrtitle

\begin{abstract}
\tolerance=3000%
Досліджено двовимірні рівноважні конфігурації директора нематичного рідкого кристалу (НРК) в комірці обмеженій двома паралельними поверхнями: планарною і просторово модульованою. Форма рельєфу модульованої поверхні описується гладкою періодичною синусоподібною функцією.
Розглянуто випадок наявності дисклінаційних ліній в об’ємі НРК строго над локальними екстремумами модульованої поверхні, що проходять вздовж гребенів і западин хвиль рельєфу.
Виникнення дисклінаційних ліній зумовлене різним типом орієнтації легкої осі директора на обмежуючих поверхнях комірки (гомеотропна орієнтація на одній з обмежуючих поверхонь і планарна на іншій) та жорстким зчепленням НРК з поверхнями.
В наближенні плоских деформацій директора отримано аналітичні вирази для розподілу директора в об’ємі комірки.
Розраховано рівноважні відстані від дисклінаційних ліній до модульованої поверхні та знайдено їх залежність від параметрів комірки (періоду та глибини хвиль рельєфу), а також від товщини комірки.
Показано, що відстані від дисклінаційних ліній до модульованої поверхні зменшуються зі зростанням глибини рельєфу.
\keywords нематичний рідкий кристал, просторово модульована поверхня,
топологічні дефекти, конформне відображення

\end{abstract}

\end{document}